# Limiting Nature of Continuum Generation in Silicon

Prakash Koonath, Daniel R. Solli and Bahram Jalali
Department of Electrical Engineering, University of California, Los Angeles
Los Angeles, CA 90095-1594
jalali@ucla.edu

**Abstract**

Spectral broadening in silicon is studied numerically as well as experimentally. Temporal dynamics of the free carriers generated during the propagation of optical pulses, through the process of two-photon absorption (TPA), affect the amplitude and phase of the optical pulses, thereby determining the nature and extent of the generated spectral continuum. Experimental results are obtained by propagating pico-second optical pulses in a silicon waveguide for intensities that span two orders of magnitude (1-150 GW/cm$^2$). These results validate the conclusions drawn from numerical simulations that the continuum generation has a self-limiting nature in silicon.

Physics and Astronomy Classification Code: 42.65 -k (Nonlinear Optics)

42.82 -in (Integrated Optics)

**Text**

       Since its experimental demonstration in 1969[1], spectral continuum generation has spawned a wide variety of research, both in terms of the approaches to generate it as well as its applications. The applications of this broad band coherent source of light are highly diverse, ranging from chemical sensing to medical imaging to high throughput telecommunication. It has been also been used in fundamental research, such as a probe for chemical reactions in a photosynthesis process, and as a tool for accurate frequency and time measurements[2,3]. Although broad continua are readily generated in optical fiber, the generation and manipulation of the continuum that lends itself easily to chip-scale integration, such as in a silicon waveguide, has a potential to impact on the various applications that utilize salient features of this radiation. The generation of spectral continuum in silicon waveguides has been explored both experimentally as well as theoretically[4-6]. A numerical study emphasizing the impact of the TPA on nonlinear phase shift at input intensity levels of ~ 13 $GW/cm^2$ was published recently[7]. These references demonstrate that the TPA reduces the optical intensity and thereby affects the continuum generation. Recently, in a demonstration of continuum carving on a silicon chip, we noted that the dynamics of the free carriers that are generated during the TPA process limit the extent of the continuum generation in a silicon waveguide[8]. In addition to the effect of TPA, the dynamics of the carriers generated by TPA have a significant impact on the spectral shape and extent of the continuum. In this context an understanding of the temporal evolution of the carrier density and its impact on both the amplitude and phase of the new frequencies generated within the pulse's time envelope is critical to assess the limits of spectral broadening in silicon. In this letter, we explore the temporal dynamics of free carrier effects numerically to evaluate their impact on spectral broadening, and validate these findings with experimental studies using optical

intensities that span more than two orders of magnitude (1-150 GW/cm$^2$). It is seen that temporal dynamics of the free carrier generation along with TPA limit the spectral broadening that may be achieved from a silicon waveguide.

In silicon, the intensity-dependent refractive index that leads to self phase modulation (SPM) and the generation of spectral continuum has two contributions: (i) the Kerr nonlinearity (ii) free-carrier refraction: the modulation of the refractive index of the medium through free carriers (electrons and holes) that are generated by TPA. It is instructive to examine these contributions in order to understand their impact on the continuum generation. The Kerr effect produces blue-shifted spectral components on the trailing edge and red-shifted spectral components on the leading edge of the optical pulse envelope. On the other hand, free-carrier refraction (FCR) causes blue-shift on both edges of the pulse envelope, shown qualitatively in figure 1a. Thus, at the leading edge of the pulse, Kerr and free-carrier effects tend to counteract each other, whereas they add at the trailing edge. The overall effect is to produce a net blue shift for the broadened spectrum. Free carriers also cause the absorption (FCA) of optical energy within the pulse, and the absorption coefficient is directly proportional to the density of the free carriers that are generated by TPA. Since free carrier density follows the time integral of the pulse shape, more carriers are generated towards the trailing edge, as the, as shown in figure 1b. Thus, the blue-shifted frequency components that reside in the trailing edge of the pulse suffer more attenuation than the red-shifted components. These two effects, namely, the counteracting nature of Kerr effect and free-carrier refraction at the leading edge of the pulse, and the significantly higher attenuation suffered by the blue-shifted frequency components in the trailing edge of the pulse, limit the amount of spectral broadening that may be obtained from a silicon waveguide.

Furthermore, TPA also reduces the peak power of the optical pulse, limiting the amount of spectral broadening. In the following, the propagation of pulses in silicon is investigated experimentally as well as numerically in order to explore the limiting nature of continuum generation in silicon.

We model continuum generation in a silicon waveguide by solving the nonlinear Schrödinger equation, which governs the propagation of the optical pulses, simultaneously with the differential equation that governs the temporal evolution of the free carriers generated by TPA [5,7,9], to study the nature of continuum generation. The free carrier absorption coefficient $\alpha_{FCA}$ and refractive index change due to free carriers $n_{FCR}$ are given by [10]:

$$n_{FCR} = -\left(8.8 \times 10^{-22} N_e + 8.5 \times 10^{-18} N_h^{0.8}\right)$$

$$\alpha_{FCA} = 8.5 \times 10^{-18} N_e + 6.0 \times 10^{-18} N_h$$

where $N_e$ and $N_h$ are the densities of electrons and holes respectively in units of cm$^{-3}$. Numerical simulations are performed by propagating transform limited 3 ps wide Gaussian pulses through a 1 cm long silicon waveguide that has a modal area of 1 μm$^2$. The measured carrier lifetime $\tau$ for such waveguides is in the range of 1-10 ns, which is much shorter than the repetition rates of 440 ns used in our measurements. Thus inter-pulse effects may safely be ignored. As waveguides with normal dispersion is considered in this study, modulation instability and solitonic effects do not play a role in the continuum generation. Four-wave mixing between newly generated spectral components is included naturally in the description of SPM through the nonlinear Schrodinger equation. In the following numerical analysis, we examine hypothetical scenarios by selectively introducing physical phenomena such as Kerr effect, free-carrier refraction and free-carrier absorption, to highlight their individual impact on spectral broadening.

Spectral broadening factor, defined as the ratio of the -20 dB spectral bandwidth at the output of the waveguide to that at the input for a transform-limited Gaussian pulse, is plotted as a function of the peak intensity of the pulse at the input of the waveguide in figure 2. The values of TPA coeffiecient $\beta_{TPA}$ and Kerr nonlinearity $n_2$ used in the simulations are shown in the figure. The dashed curve shows the spectral broadening in the presence of only Kerr effect and TPA. It is seen that, with 200 GW/cm$^2$ of peak intensity at the input, the spectrum may be broadened to around 26 nm at the output, corresponding to a broadening factor of ~ 9. Combining the Kerr effect with free carrier refractive index modulation (FCR) leads to broadening factors as high as 27 at 200 GW/cm$^2$, as depicted by the dotted curve. This is due to the fact that free carrier refraction causes a strong blue shift to the spectrum, with these spectral components residing in the trailing edge of the optical pulse. Thus, even though TPA reduces the peak power of the pulse as it propagates along the waveguide, it is still possible to obtain large broadening factors if the absorption due to free carriers was absent.

The complete picture however is obtained when the effect of free carrier absorption (FCA) is also is taken into account, as shown by the continuous curve in fig.2. Even though free carrier refraction produces a strong blue shift, these spectral components suffer higher optical losses due to the fact that free carrier absorption is higher towards the trailing edge of the optical pulse. This limits the spectral broadening factors to around 12 at the end of the 1 cm long waveguide. Figure 3 shows the optical spectra (200 GW/cm$^2$ input intensity) corresponding to the different scenarios illustrated in figure 2. It is also important to look at the influence of TPA in limiting the power available to generate continuum. The effect of optical limiting by TPA, for three different input intensities, spanning an order of magnitude, is depicted in figure 4a. After 3 mm of propagation, the peak intensities of the pulses are within 1.3 dB of each other, and most of the spectral broadening occurs in these first few millimeters as shown by figure 4b.

We have also experimentally studied the nature of continuum generation in silicon waveguides with normal dispersion. Optical pulses of ~3.5 ps duration and -3 dB bandwidth of ~2 nm were coupled to a silicon waveguide 2.3 cm in length, with a modal area of around 2.8 $\mu m^2$. Measured output spectra at 3 different input peak intensity levels, along with the spectrum of the input signal, is shown in figure 5. The self-limiting nature of continuum generation becomes apparent by comparing the spectral spread in these plots. The measured spectral broadening factor as a function of the optical intensity inside the waveguide is plotted in figure 6. A clear saturation of the broadening factor is observed experimentally, as predicted by numerical simulations. It is seen that experimental curve saturates faster than that predicted by numerical simulations. However, even at optical intensities as high as 150 $GW/cm^2$, the agreement between simulations and experimental results is within a factor of 1.3.

In the simulations, the magnitude of the free carrier induced index change is obtained through an empirical model proposed by Soref [10]. This model was developed by fitting the experimental data on refractive index change for varying levels of impurity doping concentration in silicon. Physically, this is different from the present situation where the carriers are generated in the medium by high power optical pulses. It is also known that optical nonlinearities saturate at very high optical intensities [11]. Thus, at high optical intensities, it may be necessary to modify the Soref model to include the saturation of nonlinearity. In addition, the mobility of free-carriers is influenced by the presence of strong electric fields [12]. Specifically, at high DC fields, carriers acquire energy from the field, leading to an increase in their scattering with lattice vibrations (phonons). This increased scattering reduces their mobility. To first order, the free-carrier absorption coefficient is inversely proportional to their mobility. Thus a reduction in mobility can further increase free-carrier absorption, thereby affect the spectral broadening. Although electric field due to an optical wave oscillates rapidly, qualitative predictions of the DC mobility model may be relevant. Fields strengths as high as $10^6$ V/cm exist inside the optical waveguide for an optical intensity of 200 GW/cm$^2$, and it is possible that this field influences the mobility of the free-carriers. In general, the simple model proposed by Soref might not be adequate to describe physical phenomena that manifest at very high optical intensities. However, given the simplicity of this model, it gives remarkably good agreement with experimental results, and provides valuable insights.

In summary, the spectral broadening of intense optical pulses in silicon waveguides with normal dispersion has been studied numerically and validated experimentally over optical intensities that span two orders of magnitude from 1 to 150 GW/cm$^2$. The temporal dynamics of free carrier generation and its impact on the phase and amplitude of the optical pulses limit the broadening that may be obtained from these waveguides. SPM due to Kerr effect and .FCR counteract each other at the leading edge of the optical pulse to limit the broadening on the red side, whereas FCA attenuate the blue components of the spectrum that are situated near the trailing edge of the pulse. The level of broadening may still be adequate for applications such as on-chip wavelength-division multiplexing for optical interconnects [13]. However, for applications that require ultra-broadband continuum radiation, spectral broadening obtained in silicon with pico-second pulses might not be the suitable choice. Even though the broadening factor may not be much larger than the present case, broader continua may be produced using ultra-short (10-100 femto-second pulses) optical pulses with high intensities as the input [14].

**Figure Captions**

Figure 1a. Qualitative depiction of the influence of Kerr effect and Free Carrier Refraction on the spectral broadening of an optical pulse.

Figure 1b. Generation of free carriers in an optical pulse. Free carriers follow the integral of the optical pulse and accumulate towards the trailing edge of the pulse.

Figure 2. Influence of various physical phenomena on spectral broadening factor plotted as a function of the input intensity of the pulse.

Figure 3. Simulated optical spectra at an input intensity of 200 GW/cm$^2$. Free carrier refraction blue-shifts the spectrum and free carrier absorption attenuates blue components of the spectrum, limiting the spectral broadening.

Figure 4a. Influence of TPA in reducing the optical intensity of the pulse for 3 different input intensities.

Figure 4b. Spectral broadening factor as a function of the length of the waveguide, for 3 different input intensities. Most of the broadening takes place in the first few millimeters of the waveguide.

Figure 5. Experimentally observed continua for various input intensities, shown along with the spectra of the input optical pulse, for a waveguide of length 2.3 cm and effective area 2.8 $\mu$m$^2$.

Figure 6. Comparison of experimentally observed spectral broadening factor with simulated values, plotted as a function of the input optical intensity.

# Figure 1a

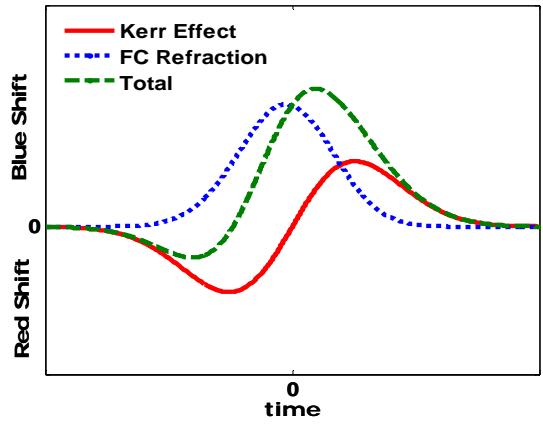

**Figure 1a. Qualitative depiction of the influence of Kerr effect and Free Carrier Refraction on the spectral broadening of an optical pulse.**

# Figure 1b

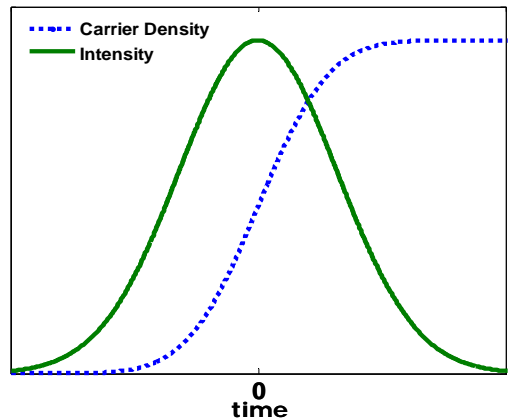

**Figure 1b. Generation of free carriers in an optical pulse. Free carriers follow the integral of the optical pulse and accumulate towards the trailing edge of the pulse.**

P. Koonath, D. R. Solli and B. Jalali

# Figure 2

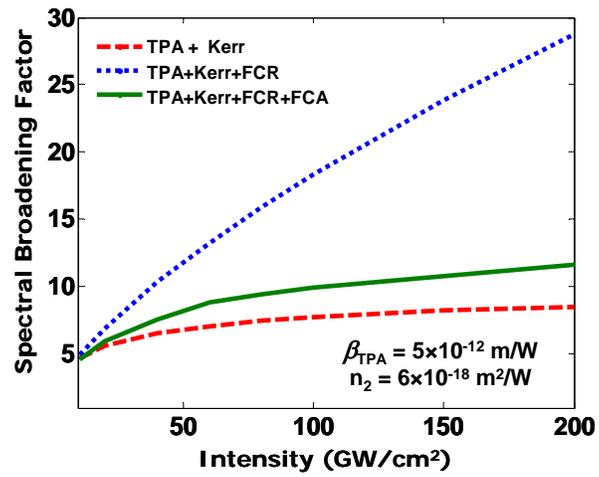

**Figure 2. Influence of various physical phenomena on spectral broadening factor plotted as a function of the input intensity of the pulse.**

P. Koonath, D. R. Solli and B. Jalali

## Figure 3

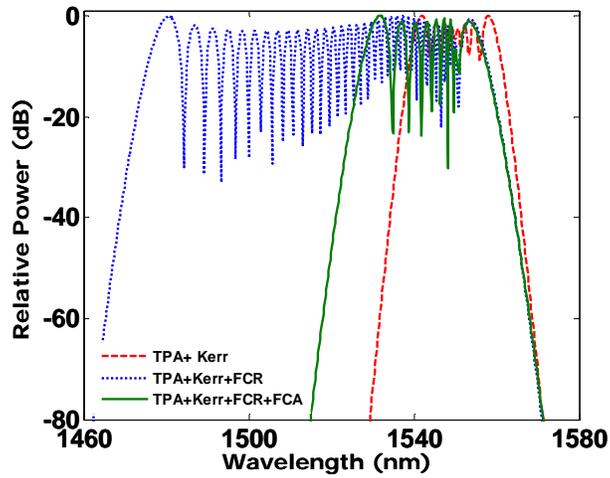

**Figure 3. Simulated optical spectra at an input intensity of 200 GW/cm$^2$. Free carrier refraction blue-shifts the spectrum and free carrier absorption attenuates blue components of the spectrum, limiting the spectral broadening.**

P. Koonath, D. R. Solli and B. Jalali

## Figure 4a

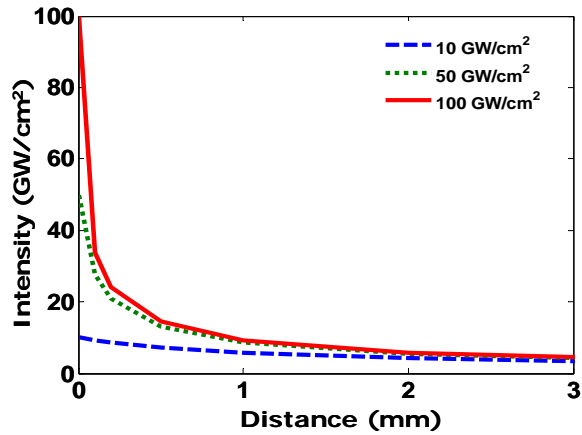

**Figure 4a. Influence of TPA in reducing the optical intensity of the pulse for 3 different input intensities.**

## Figure 4b

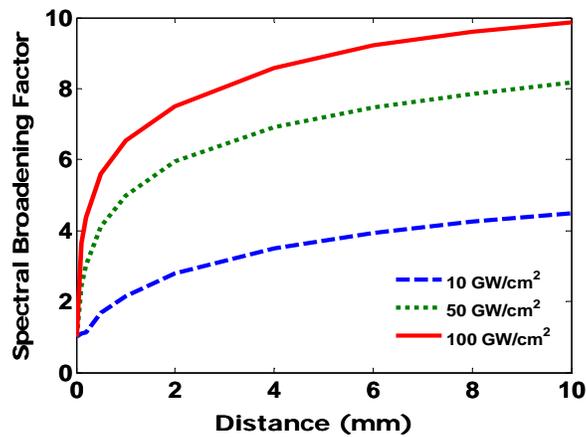

**Figure 4b. Spectral broadening factor as a function of the length of the waveguide, for 3 different input intensities. Most of the broadening takes place in the first few millimeters of the waveguide.**

P. Koonath, D. R. Solli and B. Jalali

# Figure 5

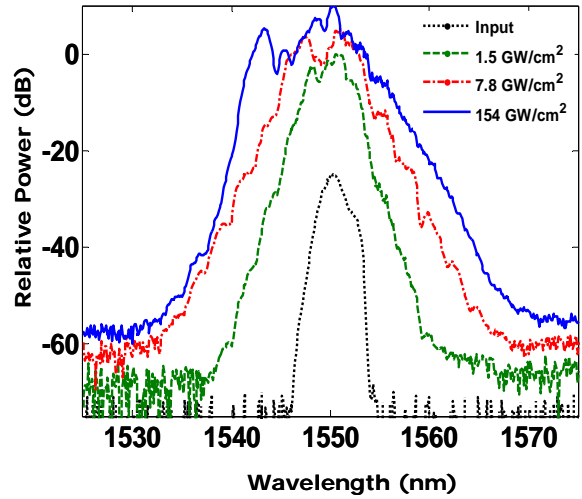

**Figure 5.** Experimentally observed continua for various input intensities, shown along with the spectra of the input optical pulse, for a waveguide of length 2.3 cm and effective area 2.8 μm².

P. Koonath, D. R. Solli and B. Jalali

## Figure 6

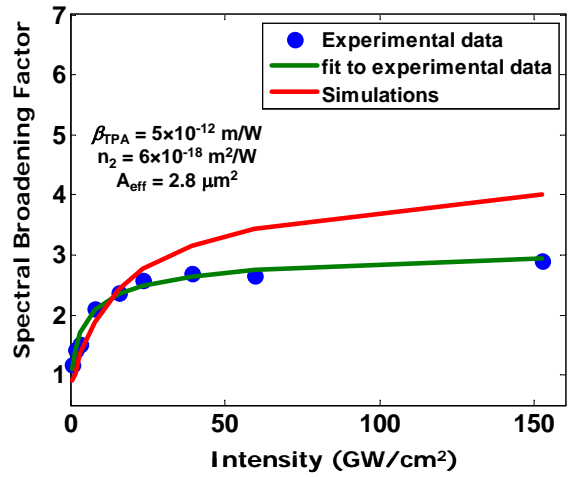

**Figure 6.** Comparison of experimentally observed spectral broadening factor with simulated values, plotted as a function of the input optical intensity.

P. Koonath, D. R. Solli and B. Jalali